\documentclass[preprint,12pt]{elsarticle0}
\usepackage[super]{cite}
\usepackage{graphicx}
\usepackage{multirow}
\usepackage{axodraw}
\usepackage{amsmath}

\def\order#1{{\mathcal O}\left(#1\right)}
\def\Li#1#2{{\mathrm{Li}}_{#1}\left(#2\right)}

\begin{document}

\begin{frontmatter}
  
\title{Forward-backward asymmetry in $e^+e^-$ annihilation into pions or kaons revisited}

\author[bltp,unidubna]{A.B. Arbuzov}
\author[unidubna]{T.V. Kopylova}
\author[bltp]{G.A. Seilkhanova}

\address[bltp]{Bogoliubov Laboratory for Theoretical Physics, JINR, Dubna, 141980 Russia}
\address[unidubna]{Department of Higher Mathematics, Dubna State University, 141980 Dubna, Russia}

\begin{abstract}
Forward-backward (charge) asymmetry in the processes of $e^+e^-$ annihilation into
a pair of charged pseudoscalar mesons is recalculated in the one-loop
approximation. The exact dependence on the meson masses is taken into account.
Results known in the literature are partially corrected.
The bulk of the charge asymmetry appears due to double-photon exchange in $s$ 
channel. Experimental studies of the asymmetry can be used to verify the 
point-like approximation used in calculation of radiative corrections due 
to emission of photons by pions or kaons.  
\end{abstract}



\end{frontmatter}

\section{Introduction and Preliminaries}	

Let us consider the process of electron-positron annihilation into
a pair of charged pions
\begin{equation}
e^+(p_+)\ +\ e^-(p_-)\ \longrightarrow\ \pi^+(q_+)\ +\ \pi^-(q_-)
\label{eq:proc}
\end{equation}
This process gives one of the most important contributions to the evaluation 
of the hadronic vacuum polarization effect extracted from 
experimental data on electron-positron annihilation 
into hadrons~\cite{Actis:2010gg}. The hadronic contribution to vacuum
polarization is then used in theoretical description of various phenomena
in particle physics, including the electron and muon anomalous magnetic
momenta, Bhabha scattering, Drell-Yan processes {\it etc.} This gives us the 
motivation to revisit the two-pion annihilation channel. The annihilation
into kaons can be considered in parallel on the same footing. These processes
are studied experimentally with high precision and modern $e^+e^-$ colliders,
see {\it e.g.} Refs.~\cite{Akhmetshin:2001ig,Fedotovich:2019znm,Kozyrev:2017agm}.

The Born level cross section of process~(\ref{eq:proc}) has the form
\begin{eqnarray}
&& \frac{d\sigma^{\mathrm{Born}}}{d c} = \frac{\alpha^2\beta^3\pi}{4s}(1-c^2)|F_\pi(s)|^2,
\qquad \beta=\sqrt{1-4m_\pi^2/s},
\nonumber \\
&& s =(p_++p_-)^2, \qquad \theta = \widehat{\boldsymbol{p}_-\boldsymbol{q}}_-, \qquad
c\equiv\cos\theta,
\end{eqnarray}
where $F_\pi(s)$ is the pion form factor, $\alpha$ is the fine structure constant,
and $m_\pi$ is the pion mass\footnote{We systematically drop terms suppressed by
the factor $m_e^2/s$.}.

\begin{figure}[ht]
\begin{picture}(400,80)(0,0)
\ArrowLine(10,70)(40,70)
\ArrowLine(40,70)(40,10)
\ArrowLine(40,10)(10,10)
\Photon(40,70)(70,70){3}{5}
\Photon(40,10)(70,10){3}{5}
\DashLine(70,70)(100,70){5}
\DashLine(70,70)(70,10){5}
\DashLine(70,10)(100,10){5}
\Text(18,65)[]{$e^{-}$}
\Text(18,19)[]{$e^{+}$}
\Text(95,65)[]{$\pi^{-}$}
\Text(95,19)[]{$\pi^{+}$}
\Text(55,0)[]{$(a)$}
\ArrowLine(120,70)(150,70)
\ArrowLine(150,70)(150,10)
\ArrowLine(150,10)(120,10)
\Photon(150,70)(180,10){2}{7}
\Photon(150,10)(180,70){-2}{7}
\DashLine(180,70)(210,70){5}
\DashLine(180,70)(180,10){5}
\DashLine(180,10)(210,10){5}
\Text(165,0)[]{$(b)$}
\ArrowLine(230,70)(260,70)
\ArrowLine(260,70)(260,10)
\ArrowLine(260,10)(230,10)
\Photon(260,70)(290,40){2}{7}
\Photon(260,10)(290,40){-2}{7}
\DashLine(290,40)(320,70){5}
\DashLine(290,40)(320,10){5}
\Text(285,0)[]{$(c)$}
\end{picture}
\caption{Box and triangle type virtual loop diagrams.\protect\label{fig_box}}
\end{figure}
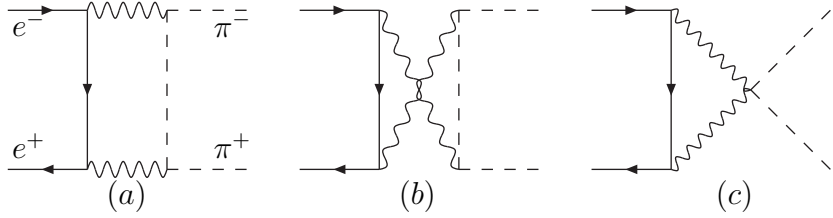

Note that the Born cross section is even in the cosine $\theta$.
But the observed cross section exhibits a certain asymmetry in $c$ which
leads to the so-called {\it charge} or {\it forward-backward asymmetry}
\begin{eqnarray}
\eta(c) = \frac{\frac{d\sigma}{d c}(c)-\frac{d\sigma}{d c}(-c)}
{\frac{d\sigma}{d c}(c)+\frac{d\sigma}{d c}(-c)}.
\end{eqnarray}
In scalar electrodynamics (sQED) with point-like pions (or kaons) this asymmetry 
comes from virtual (loop) and bremsstrahlung corrections.
So studies of this effect potentially allow to verify the applicability
of sQED to this process and therefore check the validity of the theoretical
approximations used in such calculations.
One can note also that in the $\order{\alpha}$ 
the asymmetry comes from a particular set of Feynman diagrams which
describe interference of photon emission from the initial and final state
particles, see Figs.~\ref{fig_real}, \ref{fig_box}.

\section{One-loop contributions}

Following the traditional technique of QED radiative correction calculations, 
we separate the one-loop contributions to the charge-odd part of the cross section
into three parts:
\begin{eqnarray}
\frac{d\sigma_{\mathrm{odd}}}{d c} = \frac{d\sigma^{\mathrm{Born}}}{d c}
\delta^{\mathrm{Virt}}_{\mathrm{odd}}(\lambda)
+ \frac{d\sigma^{\mathrm{Born}}}{d c}\delta^{\mathrm{Soft}}_{\mathrm{odd}}(\lambda,\Delta)
+ \frac{d\sigma^{\mathrm{Hard}}_{\mathrm{odd}}}{d c}(\Delta)\, ,
\end{eqnarray}
where $\lambda$ is a fictitious photon mass which is used to regularize
infrared singularities, $\lambda \ll m_e$. Parameter $\Delta$ defines 
separation of soft and hard photons: a photon with energy below 
$\Delta\sqrt{s}/2$ is called ``soft'' (and ``hard'' above), 
$\Delta \ll 1$.    

If the final state mesons are treated as point-like particles, 
one-loop calculations are performed in a straightforward way in sQED.
The only complication is due to the necessity to keep the exact 
dependence on the meson mass since we are interested in a wide energy range
including the threshold region.

\subsection{Soft photon emission contribution}

We take the $\order{\alpha}$ result for charge-odd soft photon contribution
from the Bremsstrahlung modules of the SANC system~\cite{Andonov:2008ga}.
It can be cast in the form 
\begin{eqnarray} \label{d_soft}
&& \delta^{\mathrm{Soft}}_{\mathrm{odd}}= \frac{\alpha}{\pi}\biggl\{
    2\ln\frac{2\Delta}{\lambda}\ln\left(\frac{1-\beta c}{1+\beta c}\right)
  - \ln^2\left(\frac{\beta(1-c)}{1-\beta c}\right)
  + 2\ln\left(\frac{\beta(1-c)}{1-\beta c}\right)
\nonumber \\ && \qquad 
  \times  \ln\left(\frac{1-\beta}{1+\beta}\right)
  + 2\ln\left(\frac{1 - 2\beta c + \beta^2}{(1-\beta c)^2}\right)
    \ln\left(\frac{\beta^2(1-c^2)}{1 - 2\beta c + \beta^2}\right)
\nonumber \\ && \qquad 
  + \ln^2\left(\frac{1 - 2\beta c + \beta^2}{(1-\beta c)^2}\right)
  + 2\Li{2}{\frac{\beta(1+c)(1-\beta)}{(1-\beta c)(1+\beta)}} 
\\ \nonumber && \qquad
  - 2\Li{2}{\frac{(1-\beta c)^2}{1-2\beta c+\beta^2}} 
  - 2\Li{2}{\frac{-(1-\beta)(1-\beta c)}{\beta(1+\beta)(1-c)}} 
  \biggr\}
  - (c \leftrightarrow -c), \qquad
\\ \nonumber &&
\Li{2}{x} = - \int_{0}^{x}\frac{dy}{y}\ln(1-y),
\end{eqnarray}
where $\Li{2}{x}$ is the dilogarithm function.

To verify the SANC analytic result for the soft photon contribution
we performed a numerical test. Numerical results received with the help of the
formulae extracted from the SANC system were compared with the corresponding
results from the direct 3-dimensional numerical integration of the relevant
part of the matrix element. The test was performed for several centre-of-mass 
energy values including points close to the threshold. 
An agreement within insignificant
uncertainties of numerical integration was observed.

In Table~\ref{ta1} we show the comparisons between results for soft
photon contributions from paper~\cite{KuraevPanov:1991aa} 
(upper lines) and the SANC system (lower lines).
One can see that a certain difference appears at energies close to the threshold
and it goes down at higher energies. One can note also that the difference is not
sensitive to the value of the soft-hard separator $\Delta$.

\begin{table}[ht]
\caption{Numerical comparison of soft photon contributions.}
{\begin{tabular}{c|ccccc} \hline\hline
$\Delta$ & $\sqrt{s}$ & $2.02 m_{\mathrm{\pi}}$ & $2.2 m_{\mathrm{\pi}}$ & $3 m_{\mathrm{\pi}}$ & $10 m_{\mathrm{\pi}}$ \\
\hline
$10^{\mathrm{-3}}$ & $\delta^{\mathrm{Soft}}_{\mathrm{odd}}$ [KP] & -0.4599587 & -0.4929014 & -0.6048826 & -0.9210629 \\ 
& $\delta^{\mathrm{Soft}}_{\mathrm{odd}}$  [SANC] & -0.4564815 & 0.4898119 & -0.6032809 & -0.9210318 \\
\hline
$10^{\mathrm{-4}}$ &$\delta^{\mathrm{Soft}}_{\mathrm{odd}}$ [KP]& -0.4045901 & -0.4336882 & -0.5328269 & -0.8164925 \\
& $\delta^{\mathrm{Soft}}_{\mathrm{odd}}$ [SANC]& -0.4011130& -0.4305987 & -0.5312254 & -0.8164613 \\
\hline
$10^{\mathrm{-5}}$ & $\delta^{\mathrm{Soft}}_{\mathrm{odd}}$ [KP] & -0.3492217 & -0.3744749 & -0.4607713 & -0.7119220\\
& $\delta^{\mathrm{Soft}}_{\mathrm{odd}}$ [SANC] & -0.3457446 & -0.3713856 & -0.4591698 & -0.7118909 \\ 
\hline\hline
\end{tabular} \label{ta1} } 
\end{table}

\subsection{Hard photon emission contribution}

\begin{figure}[ht]
\begin{picture}(400,160)(0,0)
\ArrowLine(10,150)(25,135)
\ArrowLine(25,135)(40,120)
\ArrowLine(40,120)(10,90)
\Photon(40,120)(70,120){3}{5}
\Photon(25,135)(40,150){3}{5}
\Vertex(70,120){3}
\DashLine(70,120)(100,150){5}
\DashLine(70,120)(100,90){5}
\Text(21,152)[]{$e^{-}$}
\Text(24,92)[]{$e^{+}$}
\Text(106,145)[]{$\pi^{-}$}
\Text(106,96)[]{$\pi^{+}$}
\Text(45,152)[]{$\gamma$}
\Text(55,87)[]{$(a)$}
\ArrowLine(120,150)(150,120)
\ArrowLine(150,120)(135,105)
\ArrowLine(135,105)(120,90)
\Photon(150,120)(180,120){3}{5}
\Photon(135,105)(150,90){3}{5}
\Vertex(180,120){3}
\DashLine(180,120)(210,150){5}
\DashLine(180,120)(210,90){5}
\Text(131,152)[]{$p_{-}$}
\Text(134,92)[]{$p_{+}$}
\Text(216,144)[]{$q_{-}$}
\Text(216,95)[]{$q_{+}$}
\Text(155,92)[]{$k$}
\Text(175,87)[]{$(b)$}
\ArrowLine(230,150)(260,120)
\ArrowLine(260,120)(230,90)
\Photon(260,120)(290,120){3}{5}
\Photon(305,135)(320,120){3}{5}
\Vertex(290,120){3}
\DashLine(290,120)(320,150){5}
\DashLine(290,120)(320,90){5}
\Text(275,87)[]{$(c)$}
\ArrowLine(60,70)(90,40)
\ArrowLine(90,40)(60,10)
\Photon(90,40)(120,40){3}{5}
\Photon(135,25)(150,40){3}{5}
\Vertex(120,40){3}
\DashLine(120,40)(150,70){5}
\DashLine(120,40)(150,10){5}
\Text(105,7)[]{$(d)$}
\ArrowLine(170,70)(200,40)
\ArrowLine(200,40)(170,10)
\Photon(200,40)(230,40){3}{5}
\Photon(230,40)(260,40){3}{5}
\Vertex(230,40){3}
\DashLine(230,40)(260,70){5}
\DashLine(230,40)(260,10){5}
\Text(215,7)[]{$(e)$}
\end{picture}
\caption{Bremsstrahlung Feynman diagrams.\protect\label{fig_real}}
\end{figure}
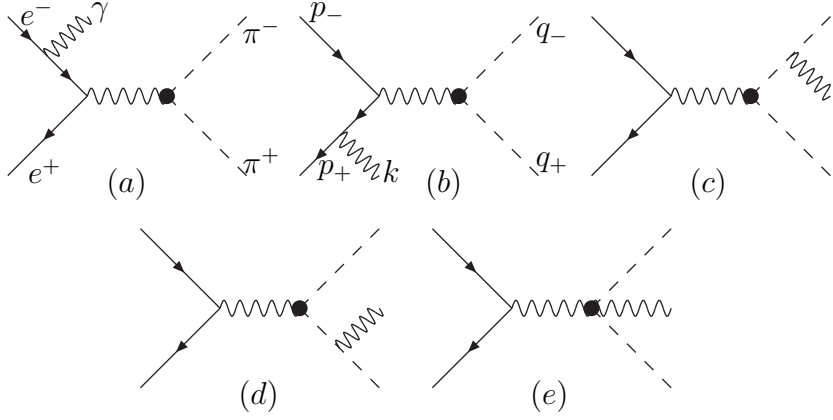

The Feynman diagrams for the process of $e^+e^-$ annihilation into
a pair of charged scalar mesons accompanied by emission of a real photon
are shown in Fig.~\ref{fig_real}. 
The filled circles in the diagrams denote the pion form factor. 
Note that we do not introduce form factors in the vertexes describing
real photon emission, {\it i.e.} we adapt the approximations of point-like
pions (or kaons) in the description of the bremsstrahlung contribution.  

Using the standard techniques of 
spinor and scalar QED, we reproduced the full analytic result for 
the differential hard photon bremsstrahlung contribution given by
Eqs.~(30) and (31) in paper~\cite{Arbuzov:2005pt}. 
The charge-odd contribution to annihilation cross sections
comes from the interference of the initial and final state radiation.
So, the interference of the Feynman amplitudes represented
in Fig.~\ref{fig_real} by diagrams $(a)$ and $(b)$ with the $(c)$,
$(d)$, and $(e)$ ones is relevant. The corresponding contribution is 
proportional to the product of pion form factors of different arguments:
$\Re\mathrm{e} (F_\pi(s)F^*_\pi(s_1)$ where $s_1 = (q_++q_-)^2=(p_++p_--k)^2$.

\subsection{Virtual loop contribution}

We compute the virtual loop diagrams in the point-like pion approximation
without introduction of meson form factors. In order to match the contribution of
virtual corrections with the other ones, we cast it in the form
\begin{eqnarray}
\frac{d\sigma_{\mathrm{odd}}^{\mathrm{Virt}}}{d c} = \frac{d\sigma^{\mathrm{Born}}}{d c}
\delta^{\mathrm{Virt}}_{\mathrm{odd}}(\lambda),
\end{eqnarray}
where the pion form factor is restored in the factorized Born level cross section.

The result of our calculations is
\begin{eqnarray} \label{d_virt}
&& \delta^{\mathrm{Virt}}_{\mathrm{odd}}(\lambda) = \frac{\alpha}{\pi}\Biggl\{
2\ln\frac{\sqrt{s}}{\lambda}\ln\left(\frac{1+\beta c}{1-\beta c}\right)
+ \frac{1}{\beta^2(1-c^2)}\biggl\{(1-\beta c) \biggl[ - l_-^2 
\nonumber \\ && \qquad 
+ 2\rho L_- + 2l_-L_- 
- 2\Li{2}{\frac{1-\beta^2}{2(1-\beta c)}} 
- \frac{(1-\beta)^2}{2\beta}\left(\frac{\rho^2}{2}+\frac{\pi^2}{6}\right)
\nonumber \\ && \qquad 
+ \frac{1+\beta^2}{\beta}\biggl(\rho\ln\frac{2}{1+\beta} 
+ \Li{2}{-\frac{1-\beta}{1+\beta}} + 2\Li{2}{\frac{1-\beta}{2}} \biggr)
\biggr]
\nonumber \\ && \qquad 
+ (1-\beta^2)\biggl( \frac{l_-^2}{2} - L_-(\rho + l_-)
+ \Li{2}{\frac{1-\beta^2}{2(1-\beta c)}} \biggr)
\biggr\}  - (c \leftrightarrow - c)
\Biggr\},
\nonumber \\ && 
l_- = \ln\frac{1-\beta c}{2}\, , \qquad
L_- = \ln\left(1 - \frac{1-\beta^2}{2(1-\beta c)}\right), \qquad
\rho = \ln\frac{s}{m^2_\pi}\, .
\end{eqnarray}

One can see that term with the logarithm of the auxiliary photon mass completely
cancels out with the corresponding term in the soft photon 
contribution~(\ref{d_soft}).
Note also that the virtual (and soft) photon contributions are free from the so-called
large logarithms $L\equiv \ln(s/m_e^2)$ in spite of the fact that such logs do appear in
individual loop integrals. Formula~(\ref{d_virt}) coincides with Eq.~(1.8) from
Ref.~\cite{KuraevPanov:1991aa} except the sign before the following dilogarithm
$\Li{2}{-\frac{1-\beta}{1+\beta}}$. One can see that the improper sign before 
this term in Ref.~\cite{KuraevPanov:1991aa} is a misprint since in the same paper
one can find the opposite sign in front of this dilogarithm in the relevant 
loop integral $F_Q$ given in Eq.~(1.5). Unfortunately the incorrect sign was
reproduced in Refs.~\cite{Arbuzov:1997je,Arbuzov:2005pt}.

\section{Discussion and Conclusions}

It is interesting to note that the odd part of the virtual contribution~(\ref{d_virt})
contains terms proportional to $1/\beta^2$. They appear due to the
forced factorization of the Born cross section. Nevertheless numerical
estimates show that this double pole behavior is completely cancelled out
so that the sum of the virtual and soft odd contributions 
$\delta^{\mathrm{Virt}}_{\mathrm{odd}}+\delta^{\mathrm{Soft}}_{\mathrm{odd}}$
behaves as $\beta$ to the first power in the threshold 
region\footnote{Before the correction of the bug in the formula for 
the virtual loop contribution, the improper $1/\beta^2$ behavior
has been apparent in numerical results.}.  
Remind that for the even part of the same cross section we have
the $1/\beta$ effect in radiative corrections at the threshold 
due to the Coulomb final state interactions~\cite{Arbuzov:2011ff}. 

In this way we revisited the charge asymmetry in the processes
of electron-positron annihilation into a pair of charged
pseudoscalar particles (pions or kaons) at low energies. Certain corrections
to the earlier calculations~\cite{KuraevPanov:1991aa} of this quantity are found.
These corrections are already implemented in the updated version of the MCGPJ
event generator~\cite{Arbuzov:2005pt}. 

There is an obvious disagreement of our result with the charge-odd 
part of the sum of virtual and soft photon contributions given by
formula~(50) in Ref.~\cite{Hoefer:2001mx}. In particular, one can see 
there a logarithmic singularity in the electron mass, which is not
common in an interference of ISR and FRS amplitudes. 

Since the asymmetry is a one-loop effect proportional to $\alpha\sim 1/137$
and since it doesn't contain large logarithms, the typical magnitude of the
effect doesn't exceed the percent level. Moreover, the asymmetry is suppressed
in the threshold region by the meson relative velocity to the first power.
Nevertheless as we noted in the introduction, experimental studies of the asymmetry 
at modern and future $e^+e^-$ colliders would be useful verify the applicability 
of the point-like point (kaon) approximation.
Note that the box-type loop amplitudes as well as the initial-final state 
interference represent the so-called double-photon exchange effect. Here it
is in the $s$ channel. Verification of different approximations to describe
such double-photon exchange processes is important for a better understanding
of off-mass-shell hadron propagator behavior. So we advise the experimental 
community to pay attention to measurements of the charge asymmetry.

\section*{Acknowledgments}

We are grateful to G.V.~Fedotovich, and F.V.~Ignatov
for critical remarks and stimulating discussions.
This work was supported by RFBR grant 20-02-00441.

\end{document}